\documentclass[aps,prl,twocolumn,superscriptaddress,showpacs]{revtex4}
\usepackage{graphicx}  
\usepackage{dcolumn}   
\usepackage{bm}        
\usepackage{amssymb}   
\usepackage{amsmath}   
\usepackage[usenames]{color}

\date{\today}
\begin{document}
\title{Cosmic Bose dark matter}

 \author{Iv\'an Rodr\'iguez-Montoya\footnote{email: rrodriguez@fis.cinvestav.mx} }
 \affiliation{Departamento de F\'{\i}sica, Centro
de Investigaci\'on y de Estudios Avanzados del I.P.N.\\
Apdo. Post. 14-740, 07000, M\'exico, D.F., M\'exico} 

\author{Abdel P\'erez-Lorenzana}
 \affiliation{Departamento de F\'{\i}sica, Centro
de Investigaci\'on y de Estudios Avanzados del I.P.N.\\
Apdo. Post. 14-740, 07000, M\'exico, D.F., M\'exico} 

\author{Eduard~De~La~Cruz-Burelo}
 \affiliation{Departamento de F\'{\i}sica, Centro
de Investigaci\'on y de Estudios Avanzados del I.P.N.\\
Apdo. Post. 14-740, 07000, M\'exico, D.F., M\'exico} 

  \author{Yannick Giraud-H\'eraud}
 \affiliation{Laboratoire AstroParticule et Cosmologie (APC), CNRS: UMR 7164,  \\
 Universit\'e Denis Diderot Paris 7. 10, rue A. Domon et L. Duquet, 75205 Paris, France}

 \author{Tonatiuh Matos}
 \affiliation{Departamento de F\'{\i}sica, Centro
de Investigaci\'on y de Estudios Avanzados del I.P.N.\\
Apdo. Post. 14-740, 07000, M\'exico, D.F., M\'exico}

\begin{abstract}
 Cold and hot dark matter (CDM, HDM) imprint distinctive effects on the cosmological observables,
 naturally, they are often thought to be made of different kinds of particles.
 However, we point out that CDM and HDM could share a common origin,
 a Bose-Einstein condensate emerging with two components.
 In this framework, the mass and temperature constraints on HDM contain fundamental information of CDM,
 but also, the critical temperature of condensation shall be larger than the HDM temperature.
 We discuss two scenarios: a gas made of bosons and a gas made of boson-antiboson pairs.
 We use some cosmological data surveys to test the idea and constrain the bosonic DM parameters,
 including a forecast for the Planck mission.
 We find that the bosonic DM picture is consistent with data in the first scenario,
 although, the second one might be increasingly interesting for future data surveys.
\end{abstract}
\pacs{98.80.Ft; 95.35.+d}

\maketitle


One of the most intriguing questions in modern cosmology comprises
the nature of the so far unidentified one quarter part of the Universe content,
known as dark matter (DM).
Cosmological observations seem to support the idea that DM
is made of some kind of non baryonic, non relativistic, and
weakly interacting massive particles.
Many efforts to elucidate this issue
have been taken over the last decade, mainly motivated by the idea that the
answer to the riddle will very likely change our current understanding of matter
and its interactions.
One favorite alternative is that DM could be sourced by a Bose-Einstein (BE) condensate
made of light particles.
Several noteworthy candidates have been addressed in this category
e.g., axions and axion-like particles \cite{AxionBEC},
scalar fields \cite{SFDM} and perhaps more exotic species such as bosonic neutrinos and massive photons \cite{exotic}.
The landscape of bosonic DM appears attractive enough for a deliberate study and test of its extents of validity.
Our pursuit is not to analyze one specific candidate, but instead, outline generic properties of bosonic DM
and explore their constraints with analysis of cosmological data.

At very early stages of the Universe,
DM is thought to be in thermal contact, either with itself or other species,
composing a \textit{primeval fireball}.
DM candidates can be classified according to their particle velocities dispersion, for our purposes:
(1) Particles with vanishing velocity dispersion are 
termed \textit{cold} dark matter (CDM).
(2) \textit{Hot} dark matter (HDM) are particles that, being ultrarelativistic at early stages,
become nonrelativistic only at recent epochs.
The standard candidate for HDM is the neutrino,
whose decoupling thermal energy is at the MeV's scale
and the temperature today is around 1.95 K.
During the expansion of the Universe, while HDM is \textit{free-streaming}, CDM is gravitationally clustering.
A Universe dominated by HDM would predict a suppression of small matter structures, in contradiction with measurements.
Hence, CDM is favored to form the overall structures and compose most of the DM in the Universe.
These two families are usually assumed to be unrelated to each other;
nevertheless, the DM paradigm does not preclude scenarios that involve jointly CDM and HDM.
Let us now introduce our assumptions framework, and call it the \textit{bosonic DM picture}:
DM is dominated by particles in a BE condensate state.
Being relativistic at decoupling from the primeval fireball,
the ground state is composed of \textit{condensed bosons} which act as CDM,
but a complementary fraction of \textit{thermal bosons} resides in excited states.
As the Universe expands the temperature decreases, then,
thermal bosons might pass through a nonrelativistic transition,
being present today as bosonic HDM.
Following this idea, we describe the full bosonic system in equilibrium as the sum of two momenta distributions:
thermal bosons obey the standard BE momenta distribution, while
that of condensed bosons exhibits a \textit{delta-like} shape.
This allows us to evolve thermal bosons in phase-space and condensed bosons as a fluid.
Bosonic DM evolves along with baryons, photons, and other species, coupled only gravitationally.
These dynamics determine the observed matter structures
and the cosmic microwave background (CMB) pattern,
which can be tested with cosmological data today.
We discuss two scenarios likely to yield relativistic BE condensation:
a gas formed purely of bosons \cite{relBos} and a gas of boson-antiboson pairs \cite{antibosons}.
We analyze both through their expressions for the critical temperature of condensation,
thereby, testing their extents of consistency as DM candidates.
Given the distinctive  effect of CDM and HDM, the bosonic DM picture might be a way
to infer information of CDM, like the mass, extracted from thermal bosons,
but also, their temperature should be related to interaction couplings of CDM particles.

Let us begin to outline the BE condensation inside the primeval fireball,
at \textit{zero}-order in perturbation theory.
The process of condensation should be driven by self interactions of high-energy
bosonic particles \cite{KineticsBEC}.
Accordingly, bosons obey the BE momenta distribution
$f_{\text{BE}}^{(0)}(\mathbf{p}) = g_{\phi} /\left( e^{(E-\mu)/T_{\phi}} - 1 \right) $
in the relativistic limit $m << T_{\phi}$,
where $m$ is the mass of bosons, $T_{\phi}$ their temperature, $\mu$ the chemical potential,
and $g_{\phi}$ is the number of internal degrees of freedom.
$f_{\text{BE}}$ defines the critical temperature of condensation $T_{c}$.
In the relativistic regime,
if the bosons, $\phi$, are their own antiparticle, then
$T_{c}^{\phi} = \left( \pi^{2} n^{(0)} / g_{\phi} \zeta(3) \right) ^{1/3}$ \cite{relBos},
where $\zeta(3)\approx 1.202$;
but, in the case of abundant boson-atiboson $\phi\bar{\phi}$ pairs production
$T_{c}^{\phi \bar{\phi}} = \left( 3 \Delta n^{(0)} / g_{\phi} m \right) ^{1/2}$\cite{antibosons}.
Here, $n^{(0)}$ is the total number density, and
$\Delta n^{(0)}$ the excess of bosons over antibosons number density.
In either case, at temperatures below the critical temperature and the limit $\mu \rightarrow m$,
the theory predicts a phase transition in which a large fraction
becomes allocated in the lowest energy occupation numbers, forming a coherent state.
Henceforth, the momenta distribution \textit{blows-up}, taking a singular shape;
when thermal equilibrium is reached, it reads
$f^{(0)}(\mathbf{p}) = n^{(0)}_{c} \delta^{3}(\mathbf{p}) + f^{(0)}_{\text{BE}}(\mathbf{p}) $
 \cite{pomeau},
where $n_{c}$ is the number density of condensed bosons.

In the instantaneous decoupling approximation,
the subsequent evolution of the full bosonic system is affected only
by the expansion rate of the Universe and small gravitational instabilities.
These instabilities are expressed as linear perturbations to the space-time geometry.
In conformal newtonian gauge, the metric reads
$ ds^{2}=a(\tau)^{2}\left[ -(1+2\psi)d\tau^{2}+(1-2\varphi)dx^{i}dx_{i}\right]$.
Here $a$ is the scale factor of the Universe's expansion,
$\tau$ the proper time,
and $\psi$ along with $\varphi$ are scalar modes.
We describe  the effects on the dynamics of thermal bosons by the usual
approach to HDM \cite{neutrinos}.
The momenta distribution is perturbed to first order as
$f_{T}(\mathbf{x},\mathbf{p},\tau) = f^{(0)}(p) \left[ 1 + \Psi(\mathbf{x},\mathbf{p},\tau) \right] $,
with $f^{(0)}(p) = g_{_{\phi}} / (e^{p/T_{\phi}} - 1) $.
Then, $\Psi$ is expanded in a Legendre series to obtain a hierarchical system of coupled Vlasov equations.
On the other hand,
the phase spaece distribution of condensed bosons may be written as a narrow Maxwell-like one \cite{BECs},
\begin{equation}
 f_{c}(\mathbf{p}) = (2\pi)^{3/2} \frac{ n_{c}(a)}{\sigma(a)^{3}} \exp{ \left(  -\frac{ (\mathbf{p} + m\mathbf{v})^{2} }{2\sigma(a)^{2}} \right)  },   \label{eq_bec_dist_func}
\end{equation}
where $\upsilon = |\mathbf{v}|$ is the linear perturbation bulk flow speed.
The width $\sigma$ of the distribution is certainly small,
manifesting the distinctive \textit{blow-up} of BE condensation.
The spatial overdensities of condensed bosons are defined in terms of $n_{c} = n_{c}^{(0)} (1 + \delta )$.
From the Vlasov equation for $f_{c}$, two fluid equations can be obtained for
$\delta$ and $\upsilon$.
In Fourier space they read
$\dot{\delta}_{k} + ik\upsilon_{k} = 3 \varphi_{k}$, and
$\dot{\upsilon}_{k} + \frac{\dot{a}}{a} \upsilon_{k} = ik\psi_{k}$ \cite{dodelson},
where $k=|\mathbf{k}|$ is the magnitude of the Fourier wave-number vector.
These are formally the dynamic equations of CDM,
often obtained assuming very massive particles.
In contrast, condensed bosons are expected to be very light;
but also, the narrow shape of $f_{c}$
makes negligible all quadratic terms of $p/E \sim p/m$ for the condensed fraction.
Hence, under this framework, even light particles could behave as CDM,
provided they posses small velocities dispersion,
which is attainable by the coherent state of condensed bosons.

The evolution of bosonic DM after decoupling is as follows.
Their velocity and temperature can be affected only by the expansion rate of the Universe.
Also, in a reference frame, comoving with the expansion of the Universe,
the total number density of bosons is conserved.
Consequently, the number density of particles evolves like $a^{-3}$ and
the temperature, as well as the critical temperature, like $a^{-1}$.
If thermal bosons become nonrelativistic at recent epochs ($m \gtrsim T_{\phi}$),
the \textit{zero}-order energy density reads
$\rho_{nr}^{(0)} = m ( n_{c}^{(0)} + n^{(0)}_{T} ) $,
where $n_{T}$ is the number density of thermal bosons and depends on their temperature.
Thus, the content of bosons has two nonrelativistic components today:
the content of bosonic CDM (B-CDM) $\Omega_{\textsc{c}}$ and bosonic HDM (B-HDM) $\Omega_{H}$,
which can be simultaneously estimated using cosmological data.
This is an interesting scenario because cosmological data can provide constraints on the mass of bosonic DM by means of
\begin{eqnarray}
 m = \frac{\Omega_{\textsc{c}} \rho_{cr} }{n^{(0)}_{c} }  =  \frac{\Omega_{H} \rho_{cr}}{n^{(0)}_{T}} , \label{eq_thermal_mass}
\end{eqnarray}
where $\rho_{cr}$ is the critical density of the Universe as measured today
in terms of the Hubble constant $H_{0}$ \cite{hubble}.

In Eq. (\ref{eq_thermal_mass}), we need the temperature in order to know the mass of bosonic DM.
We can obtain it by recalling that the Universe expands adiabatically,
i.e., the entropy remains constant in a comoving reference frame.
At very early epochs, when the Universe is dominated by radiation, the entropy is
$ s = (2\pi^{2} / 45 ) \, g_{*}(T) \, T^{3} $ \cite{kolb},
where $T$ is the temperature of the primeval fireball,
and $g_{*}$ is an effective number of internal degrees of freedom
of all the relativistic species in thermal contact.
(CMB photons are the last species to decouple,
thus, today $g_{*,0}=2$, corresponding to two photon polarization states.)
The entropy leading order term of the bosonic gas is
$ s_{\phi} \sim T_{\phi}^{3}$,
as a consequence, the temperature of thermal bosons is proportional to that of CMB photons.
This proportion depends on the number of relativistic degrees of freedom
of the thermal bath from which they decouple,
which could include standard and even some exotic species.
To ilustrate, after neutrino decoupling ($T \lesssim $ 1 MeV),
electrons and positrons $e^{\pm}$ start to annihilate into photons,
leading to an increase in the temperature of photons with respect to neutrinos
by a factor proportional to the internal degrees of freedom of electrons and positrons.
Consequently, the temperature of neutrinos today
(and thereby, the usual temperature of HDM) is
$ T_{\nu,0} = \left( 4/11 \right) ^{1/3} T_{\text{cmb}} $,
where $T_{\text{cmb}}$ is the measured temperature of the CMB.
For bosons, we model any deviation from the usual HDM temperature by means of
\begin{equation}
 T_{\phi,0} = \left( \frac{11}{4} + g_{\text{x}} \right) ^{-1/3} T_{\text{cmb}}, \label{eq_gexotic_temp}
\end{equation}
where $g_{\text{x}}$ is negative (positive) if bosons decouple during (before)
the process of $e^{\pm}$ annihilations.
A particular model of bosonic DM in this scheme would contain a relation of
$g_{\text{x}}$ with their interaction couplings,
here, we regard $g_{\text{x}}$ just as a free parameter.
An increase of $g_{\text{x}}$ reduces the free-streaming of bosonic HDM,
yielding less suppression of small scale matter structures, 
resembling the effect of CDM.

As an additional remark, we address a bound on the strength of bosons self interactions.
It comes from the fact that our study relies on the range of validity of the diluteness approximation \cite{BECs},
which can be written for condensed bosons in terms of the scattering cross section,
$\sigma_{scatt} \ll (n^{(0)}_{c})^{-2/3}$.
Besides, as a minimal condition, the content of B-CDM should be at least larger than baryonic; we find
$ \sigma_{scatt} \ll 8\pi  ( m / \rho_{cr} \Omega_{b} ) ^{2/3} $,
which is more restrictive for ultralight models.
Conversely, for B-HDM it may be written as
$\sigma_{scatt} \ll 3/4 \; \left[ \left( 11/4 + g_{\text{x}} \right) \Omega_{H} / (g_{\phi}  \Omega_{b} ) \right] ^{2/3} \, \text{cm} ^{2}$.

In the generic study we are presenting here,
the bosonic DM parameters are addressed as the mass, $m$, and the factor $g_{\text{x}}$;
bounds on their values can be obtained from a statistical analysis on cosmological data.
To this end, we perfom a simultaneous $\chi^{2}$ analysis \cite{root}
to data provided by the 7-year WMAP, ACBAR and QUaD surveys for the CMB temperature anisotropies \cite{cmb-data},
together with the SDSS large redshift galaxy survey \cite{mps-data}.
We will call this set W+AQS.
The input model fit to data is obtained from an adapted version of the public code \textsc{camb} \cite{camb},
which includes our bosonic DM component.
We study two models of scalars ($g_{\phi}=1$) given by the case of a gas of bosons $\phi$ and boson-antiboson $\phi \bar{\phi}$ pairs;
in the relativistic limit, they differ only by a factor of two in the momenta distribution of thermal bosons.
For simplicity, we consider a flat geometry with cosmological constant
and choose to vary the following set of cosmological parameters:
the content of baryons $\Omega_{b}$,
the content of CDM $\Omega_{\textsc{c}}$,
the scalar inital amplitude $A_{s}$,
the spectral index $n_{s}$,
the optical depth of reionization $\tau_{r}$,
and one free bias parameter for the SDSS data.
We make use of flat priors with starting points at $g_{\text{x}}=0$, $m=0$,
and the best-fit values found by the WMAP team \cite{komatsu} for the rest of parameters.

\begin{figure}[t]
 \centering
 \includegraphics[width=8.0cm]{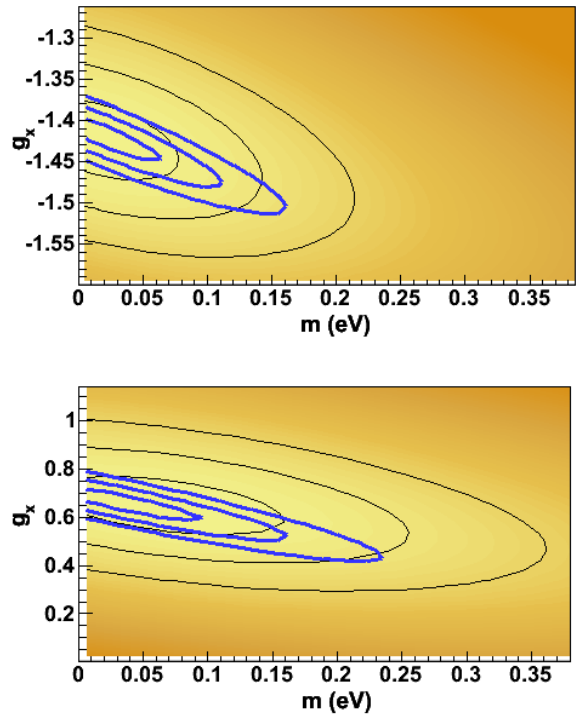}
 \caption{1,2,3-$\sigma$ probability contours from WMAP+AQS for the bosonic HDM parameters in
 the case of bosons (up) and boson-antiboson pairs (down) models. Blue contours refer to Planck+AQS.}
\label{fig:a}
 \end{figure}

\begin{table}[b]
\begin{center}
\begin{tabular}{cccccc}\hline\hline 
(95~\%)          &                         &  &  $g_{\text{x}}$   &  m (eV)         & $\Omega_{\textsc{c}}\times 10^{-1}$  \\ 
\hline
 $\phi$          & \textsc{w}+\textsc{aqs} &  & -1.42$\pm$0.11    & $\lesssim$0.14  & 2.1$\pm$0.025  \\
                 & \textsc{p}+\textsc{aqs} &  & -1.42$\pm$0.05    & $\lesssim$0.11  & 2.1$\pm$0.010  \\
  \\
$\phi\bar{\phi}$ & \textsc{w}+\textsc{aqs} &  & 0.67$\pm$0.26     & $\lesssim$0.26  & 2.5$\pm$0.030     \\
                 & \textsc{p}+\textsc{aqs} &  & 0.67$\pm$0.16     & $\lesssim$0.16  & 2.5$\pm$0.012     \\
\hline\hline
\end{tabular}
\end{center}
\caption{Posterior probabilities of bosonic DM parameters from the two data sets considered,
         for the case of bosons $\phi$ and boson-antiboson $\phi\bar{\phi}$.}
\label{tabla}
\end{table}

\begin{figure} 
 \centering
 \begin{tabular}{c}
 \includegraphics[width=8.6cm]{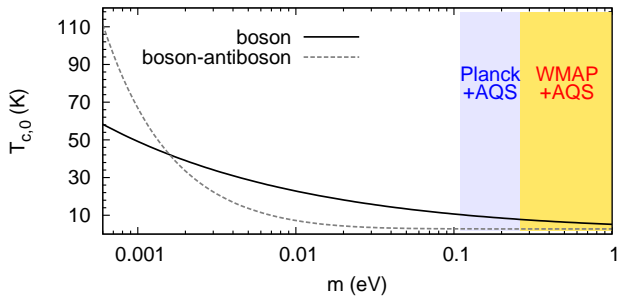}
\end{tabular}
 \caption{Critical temperatures of condensation (evaluated today) as a function of the mass for each model.
          For bosons, the lower limit $T_{c,0}^{\phi}\gtrsim 10.33$ K is higher than our best fit of $T_{0}^{\phi}=2.14$ K.
          For boson-antiboson, the limiting-critical temperature is comparable with the bosonic temperature.}
\label{fig:c}
\end{figure}

The posterior probabilities of bosonic DM parameters are ploted in Fig. \ref{fig:a} and
listed in table \ref{tabla}, from which follows constraints in the temperature
of bosons $T_{0}^{\phi}$ = 2.14$\pm$0.02 K, and
boson-antiboson gas $T_{0}^{\phi \bar{\phi}}$ = 1.91$\pm$0.05 K (95\%).
The bound in the cross section of self interactions becomes
$\sigma_{scatt} \ll$ 0.18 cm$^{2}$.
From Eq.~(\ref{eq_thermal_mass}),
we may set lower limits on the critical temperature of condensation (evaluated today):
$T_{c,0}^{\phi} \gtrsim$ 9.53 K and $T_{c,0}^{\phi \bar{\phi}} \gtrsim$ 1.91 K,
for the two cases.
It is interesting to explore the constraints that the Planck mission will achieve on our model parameters.
We take a fiducial power amplitude given by the best fit to the WMAP data,
we neglect systematic effects and assume a sky coverage of 65\%.
The data forecast is based on the contributions from cosmic variance and
noise specifications of the first three channels of the high frequency instrument (see \cite{planck});
we call this set of data as P+AQS.
As appreciated in Fig.~\ref{fig:a} and table \ref{tabla},
our Planck forecast reduces considerably the probability parameter space.
Subsequently, we find $T_{0}^{\phi}$ = 2.14$\pm$0.009 K,
$T_{c,0}^{\phi} \gtrsim$10.33 K
for bosons and
$T_{0}^{\phi \bar{\phi}}$ = 1.91$\pm$0.009 K,
$T_{c,0}^{\phi \bar{\phi}} \gtrsim$ 1.92 K
for boson-antibon,
the bound on the self interaction cross section becomes
$\sigma_{scatt} \ll$ 0.15 cm$^{2}$ (95\%).
Our result $T_{0}^{\phi} \lesssim 0.22 \, T_{c,0}^{\phi}$ indicates
consistency with the assumptions of the bosonic DM picture.
Even if the case of boson-antiboson is not conclusive in this respect,
note however, that the critical temperature depends mainly on the mass constraints.
We show this dependence in Fig.~\ref{fig:c} with the temperature fixed to our best fit.
A boson-antiboson gas exhibits a huge critical temperature if their particles are ultralight.
Therefore, better constraints on the mass of bosons by larger surveys
could yield upper bounds on the critical temperature,
perhaps in consistency with the hypothesis of BE condensation.

Finally, from a similar analysis for fermionic HDM (F-HDM) we find bounds
in the sum of neutrino masses and the number of extra relativistic species,
$\sum m_{\nu} \lesssim$0.45 eV, $N$=1.10$\pm$0.18 (95\%),
in concordance with some previous reports \cite{hannestad}.
When we compare the F-HDM and B-HDM spectra,
differences are roughly apparent at the ten percent level;
however, our $\chi^{2}$-study shows no signficant difference between B-HDM and F-HDM in their ability to fit the presented data.
This should be considered as a source of systematic uncertainty in
cosmic neutrino parameter estimations \cite{hannestad}.

In summary, we have presented a generic study of DM based on BE condensation,
from which, CDM and HDM, are intrinsically related.
From our statistical analysis, we constrain the mass of bosonic DM
(which in our framework is the same for both CDM and HDM),
the temperature, and the self interaction cross-section.
We find that the bosonic DM picture is compatible with the data used here for a gas of bosons,
and discuss how future constraints on the mass
would yield much higher bounds on the critical temperature of a boson-antiboson gas.
Our forecast for Planck data predicts important improvements in the parameters estimation.
Additionally, we observe that thermal bosons and standard neutrinos are both compatible with the presented data.
This systematic along with the properties of bosonic DM could be further studied with future and larger cosmological surveys.

We thank Massimiliano Lattanzi for helpful correspondence.
IRM thanks the kind hospitality of the
APC, Paris Diderot University,
where a stage of this work was developed.
This work was supported in part by CONACyT grants: 132061 and 106282-F.


\end{document}